# High-order sideband generation in bulk GaAs

B. Zaks, H. Banks and M. S. Sherwin

*Department of Physics and the Institute for Terahertz Science and Technology, University of California at Santa Barbara, Santa Barbara, CA 93106*

**Abstract:** When an intense THz field at frequency $f_{\mathrm{THz}}$ is applied to excitons resonantly created in bulk GaAs by a near IR laser at frequency $f_{\mathrm{NIR}}$, sidebands are observed at frequencies $f_{sideband} = f_{\mathrm{NIR}} + 2nf_{\mathrm{THz}}$, where $n$ is an integer. At temperature T=10 K, sidebands of order $-4 \leq 2n \leq 16$ are observed. Sidebands up to $10^{th}$ order persist at 170 K.

Atomic and molecular systems in the presence of intense electromagnetic fields have been of interest for decades. In particular, the investigation of atomic and molecular systems driven by extreme optical and infrared fields has led to the discovery of high-order harmonic generation (HHG)[1]. This phenomenon has been successfully modeled by an electron tunneling out of an atom, accelerating in the intense optical field and recolliding with the atomic core it left behind[2]. The development of this phenomenon has led to significant advances in attosecond technology[3] and the ability to tomographically image atomic and molecular orbitals[4]. Recent observation of HHG in a solid state system[5] could lead to the development of similar technologies for condensed matter systems.

When intense THz radiation is applied to a semiconductor, the strong ac fields can drastically alter its optical properties near the band gap. Electro-optic effects such as the dynamical Franz-Keldysh[6] and Autler-Townes[7] effects have been observed as changes to the absorption spectra, while the THz-induced generation of new frequencies of light is observed as sidebands[8,9]. Phenomena such as sideband generation and the Autler-Townes effect have required investigation in semiconductor quantum wells due to their strong formation of excitons, bound electron-hole pairs. The recent observation of high-order sideband generation (HSG) from electron-hole recollisions[10] in a quantum well presents a number of physically intriguing and technologically relevant experimental opportunities. Recollisions have been predicted in any material which supports excitons.[11,12]

In this letter, we present our observation of high-order sideband generation in bulk GaAs. Similar to previous observations in InGaAs quantum wells[10], when excitons created by a NIR laser in bulk GaAs are driven with an intense THz field, recollisions between the electrons and holes lead to the observation of high-order sidebands. The sidebands observed are separated from the NIR frequency $f_{\mathrm{NIR}}$ by multiples of twice the THz frequency $f_{\mathrm{THz}}$. The frequency of the sideband of order $2n$ is therefore given by $f_{2n} = f_{\mathrm{NIR}} + 2nf_{\mathrm{THz}}$. Consistent with observations in quantum wells, the intensity of the sidebands decays only weakly as the order increases, particularly for the highest order sidebands.

The sample investigated was a ~10 µm thick wafer of bulk GaAs. The GaAs sample was prepared from a 350 µm semi-insulating GaAs substrate wafer which was lapped down to ~100 µm and then polished with a ~0.5 µm grit polishing paper. After thinning the sample down to ~100 µm, a thick layer of photoresist was deposited and a ~1x1.5 cm rectangular hole was developed. This rectangle on the sample was then etched down to ~10 µm using a $BCl_3/Cl_2/Ar$ dry etch in a Panasonic reactive ion etcher (ICP). The sample was mounted in a closed cycle refrigerator with Apiezon N vacuum grease and cooled to temperatures as low as 10 K.

The experimental procedures were similar to those previously described in Ref. 10. The NIR laser was sent through the etched region of the sample and the intense THz radiation was focused such that the NIR and THz were co-propagating in the sample. The NIR and THz beams were co-polarized in all experiments presented. The strong THz radiation was provided by the UC Santa Barbara Free Electron Laser (FEL) and the NIR light was created by a continuous-wave titanium sapphire laser.

After sidebands were generated in the sample, they were sent to a SPEX 1403 0.85 m double spectrometer and a Hamamatsu photomultiplier tube (PMT). The lowest order sidebands were measured as a voltage from the PMT while higher order sidebands were measured by counting the number of photons incident on the PMT using an SRS SR400 gated photon counter. Due to the limited dynamic range of our detection scheme, the bias (and hence the PMT gain) was adjusted to avoid saturation of the detector when the strongest sidebands were measured. When sidebands were measured by photon counting, the measurements were taken for 20 FEL pulses and ~1 false count per 100 pulses was observed. In order to plot the data taken as a voltage on the same figure as the data taken with the photon counter, we multiplied the measured voltages by a scaling factor. The scaling factor is taken to be the ratio of the magnitude of the sixth-order ($n$ = 3) sideband as measured by photon counting compared to as a voltage. The same scaling factor is applied for all data measured as a voltage. All error bars shown are representative of the standard error of the mean.

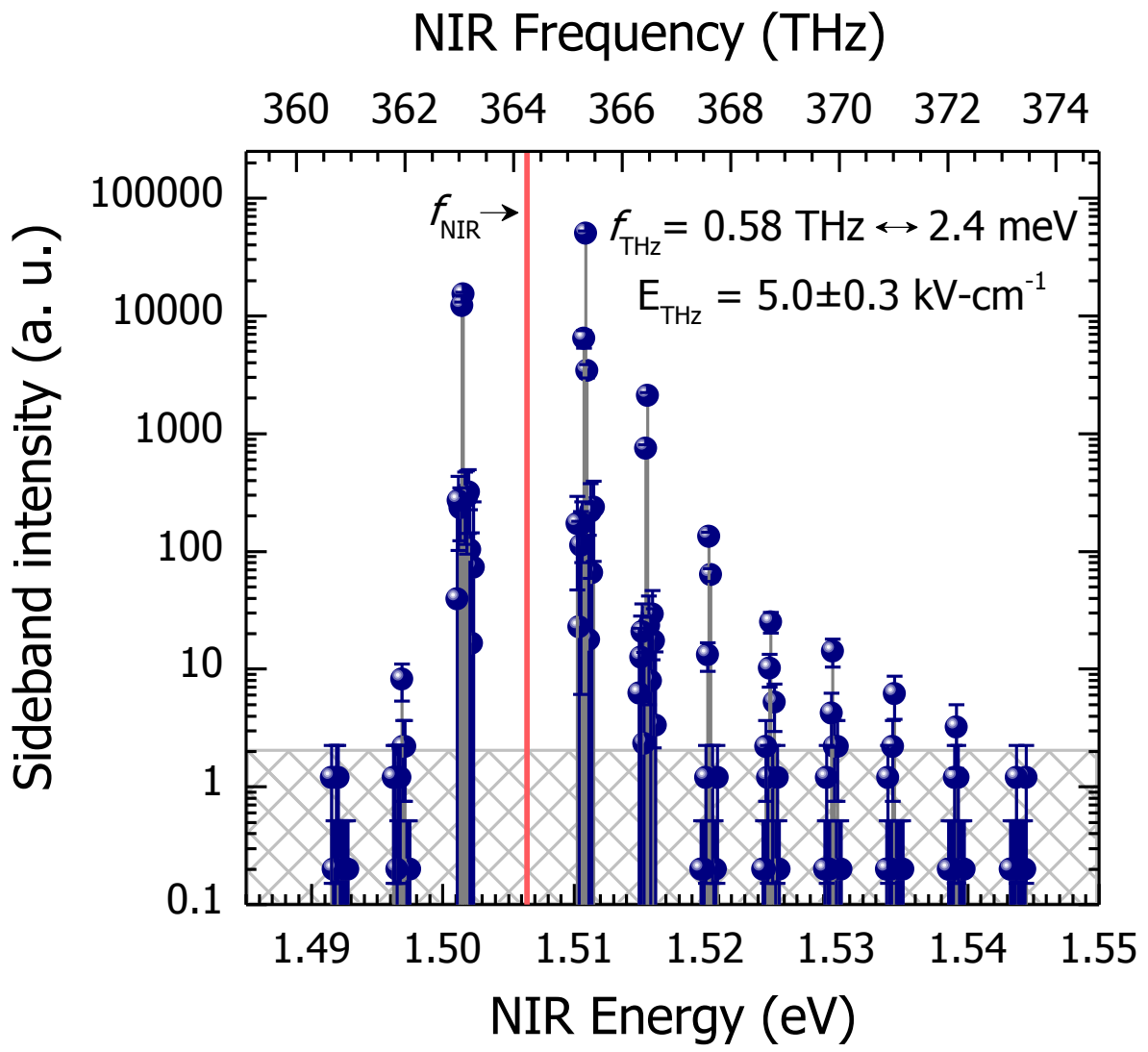

FIG. 1. (Color online) High-order sideband spectrum from a ~10 μm thick piece of bulk GaAs held at a temperature of 10K. The frequency of the intense field was 0.58 THz and the peak electric field was 5 kV-cm$^{-1}$. The grey crosshatch pattern represents the area that is below the confidence limit for identifying a sideband.

The NIR spectrum when a strong THz field was applied to the GaAs quantum well excitons is shown in Fig. 1. With a THz field of 5 kV-cm$^{-1}$ applied, sidebands of up to 14$^{th}$ ($n$ = 7) order were observed. The baseline value of 0.2 is representative of the noise level in the experiment and is plotted as the signal amplitude when no photons were measured. Any measured value with an error bar below the noise floor is considered too weak to confidently identify as a sideband, and the values below this confidence limit are represented by the grey crosshatch area in the figure. We note that 5 kV-cm$^{-1}$ is roughly half of the field required to observe a similar number of high order sidebands in a quantum well[10] (~10 kV-cm$^{-1}$). This may arise from the fact that bulk excitons are more weakly-bound than excitons in quantum wells and thus can tunnel ionize in a weaker electric field.

Similar to what is observed for HSG in quantum wells, the intensity of the highest order sidebands decays weakly with increasing order *n*. Though the ~10 µm thick bulk GaAs sample investigated here has orders of magnitude more active region than was present in the 15 nm QW sample (150 nm active region from 10 QWs), the magnitude of the signal observed was similar in both the bulk and QW experiments. According to the manuscript by Yan[12], the increased Coulomb energy of the exciton due to the confinement in the quantum wells may increase the sideband generation efficiency. This increase in efficiency may be responsible for the comparatively strong signal from thin QW sample compared to the bulk GaAs sample. Alternatively, it may be that the sidebands we have observed are near saturation and it is not possible to produce sidebands of significantly higher intensity. It should be noted that the highest sidebands observed in this experiment are well above the band gap and higher sidebands that are not observed may have been reabsorbed while propagating through the material. More investigation is necessary to identify the optimal thickness sample to generate the strongest sidebands while minimizing sideband re-absorption above the band edge.

Sidebands were only observed when the frequency of the near-IR laser was close to the onset of excitonic absorption. Figure 2 shows the dependence of sideband intensity on NIR frequency for both the positive and negative second order sidebands (Fig. 2a, $n = \pm 1$), as well as the fourth and sixth order sidebands (Fig. 2b, $n$ = 2 and 3, respectively). The intensity of all of the sidebands investigated peaked when the NIR laser had a photon energy of ~1.505 eV. Because the sideband intensity decreased quickly as the laser was detuned from this frequency and because this is the approximate energy where we expect to observe excitonic absorption, we identify this frequency as the exciton resonance. We note that this frequency differs from where we observed the exciton peak (1.510 eV) in the NIR absorption spectrum in the absence of an intense THz field (pink shaded curve). This difference in the observed exciton frequency may be due to shifts induced by the strong THz field, but more investigation will be required to understand the nature of this difference.

All of the sidebands investigated had an intensity which peaked at a NIR energy of ~1.505 eV. However, for the second order sidebands, a larger peak of the sideband intensity occurred slightly below this energy for the positive ($n = 1$, solid navy line) sideband and slightly above this energy for the negative ($n = -1$, dashed red line) sideband (Fig. 2a). The separation between the 1.505 eV peak and the larger peak for each sideband is ~5 meV, approximately twice the THz photon energy. This implies that the sideband intensity for the second order sidebands ($n = \pm 1$) is greatest when the sideband, not the NIR laser, is resonant with the exciton frequency (Fig. 2a, energy diagrams 1 and 3). Similar enhancements have been observed previously[9].

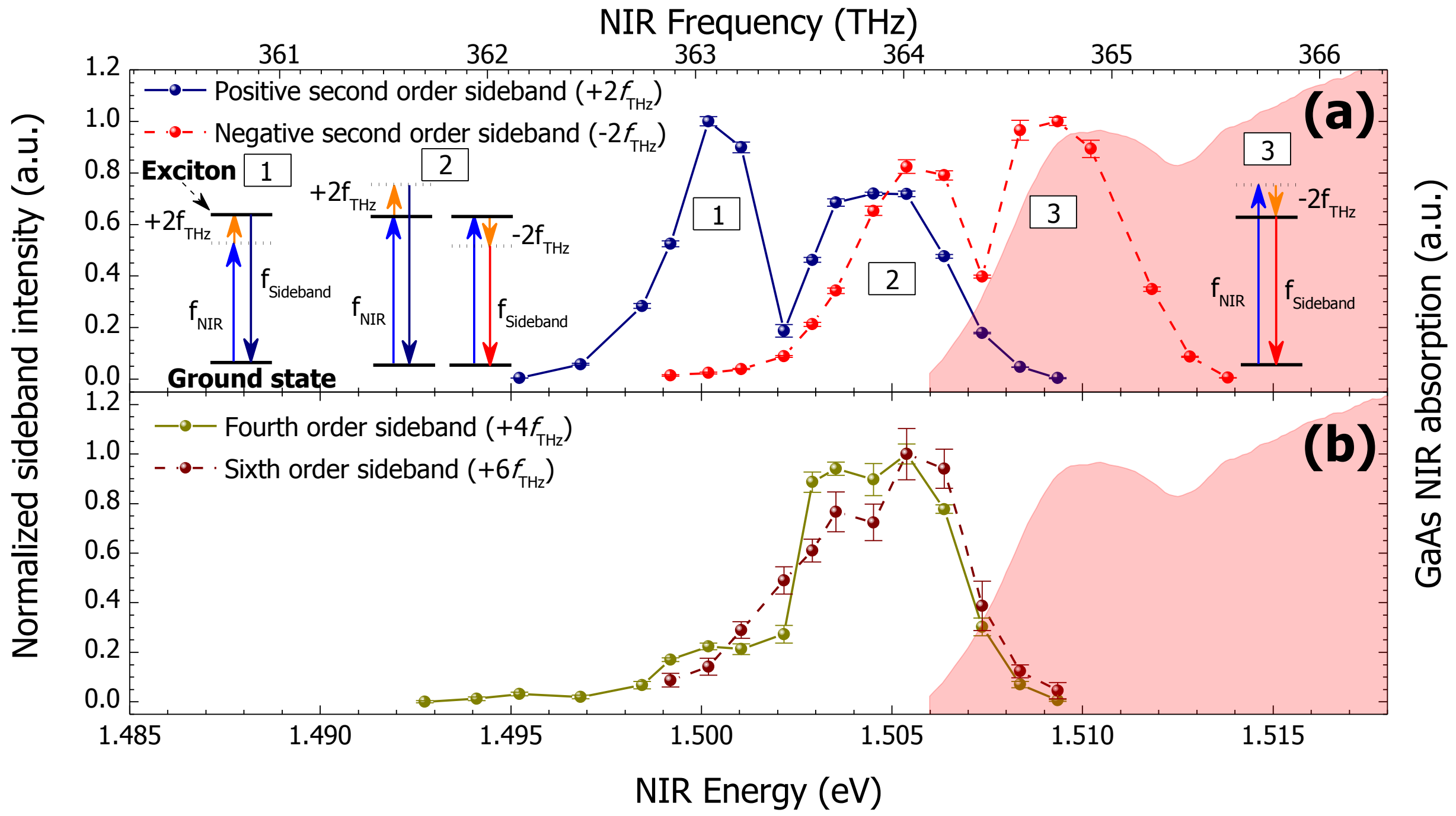

FIG. 2. (Color online) Dependence of sideband intensity on NIR laser frequency in a GaAs sample held at 10K. (a) NIR frequency dependence of the sideband intensity for the positive and negative second order sidebands ($n = \pm 1$). The sideband intensity for both sidebands peaks when the NIR laser is at 1.505 eV and we identify this energy as the exciton resonance. An energy level representation of the sidebands generated at this NIR frequency is shown in energy diagram 2. However, we observe that the sideband intensity is enhanced if the NIR laser is tuned so that the sideband frequency, which is the NIR laser plus (or minus) twice the THz frequency, is resonant with the exciton. This is shown in energy level representation for the positive 2$^{nd}$ order sideband ($n$ = +1, energy diagram 1) as well as the negative 2$^{nd}$ order sideband ($n$ = -1, energy diagram 3). In these diagrams the solid line represents the exciton energy state and the different color arrows represent the NIR and THz lasers and the sidebands generated. (b) Near infrared frequency dependence of the sideband intensity for the fourth and sixth order sidebands ($n$ = 2 and 3, respectively). The sideband intensity is greatest when the NIR laser is at ~1.505 eV. The pink shaded region on both of the graphs is the NIR absorption of the thin GaAs sample in the absence of a THz field while at 10K. The small peak at 1.510 eV is taken to be the exciton absorption in the GaAs sample.

The NIR frequency dependence of the fourth and sixth order ($n$ = 2 and 3, respectively) sidebands was also investigated (Fig. 2b). The sideband intensity for these higher order sidebands quickly decreased as the NIR frequency was detuned from the exciton resonance. There was no enhancement observed when the fourth or sixth order sidebands were resonant with the undriven exciton line.

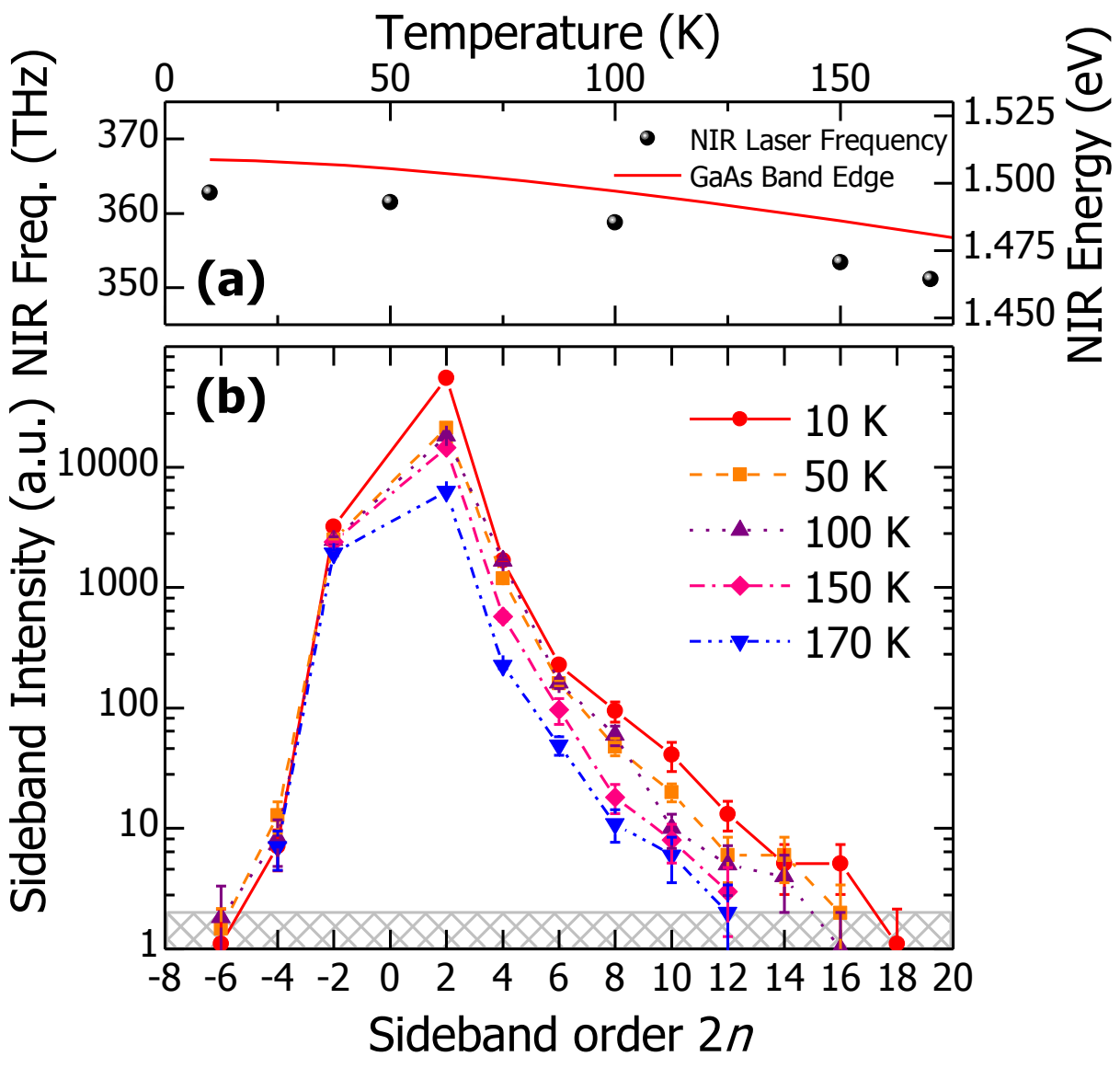

FIG. 3. (Color online) Graph depicting the dependence of sideband intensity on temperature in bulk GaAs. (a) Plot of the NIR laser frequency applied for high-order sideband generation as a function of temperature (black dots). The NIR frequency was adjusted to account for the red shift of the GaAs band edge, which is approximated by the solid red line (from http://www.ioffe.ru). Our NIR laser could not be tuned lower than ~350 THz. (b) A plot of the peak sideband intensity of each sideband as a function of temperature between 10K and 170K. Data was taken with the FEL at 0.58 THz and an electric field of ~11 kV/cm. Sidebands are identified by their order as opposed to their frequency for easier comparison between different temperatures. Sidebands of 10$^{th}$ order or greater were observed up to 170K.

Temperature dependent measurements of the sideband generation were performed between 10 K and 170 K and are shown in Fig. 3. At 170 K, the tenth order sideband was still detected. In order to create excitons at different temperatures, the frequency of the NIR laser was adjusted to track the temperature-dependent shift of the GaAs band edge. A graph of the NIR frequency applied at different temperatures is shown Fig. 3a. Figure 3b shows the peak sideband intensity for each sideband observed for temperatures between 10 K and 170 K. The NIR frequency scale is removed and the sideband intensity is plotted as a function of sideband order 2*n* to simplify the comparison of data taken at different temperatures. Due to the limited tuning range of the NIR laser, the exciton frequency could not be addressed at temperatures above 170 K, and it is likely that sidebands persist at higher temperatures. The sidebands plotted in Fig. 3 were taken with an applied THz field of ~11 kV-cm$^{-1}$. At 10 K, sidebands of up to 16$^{th}$ order were observed. The electric field applied in these measurements is greater than the electric field that was applied to take the data presented in Fig. 1 and was achieved by use of the FEL cavity-dump coupler[13].

The most sensitive measurements made in this experiment were performed with a photon counter. Because the cavity-dumped pulse was only 40 ns and the non-cavity dumped FEL pulse was ~1 μs, the measurement time for photon counting without the cavity dump was ~25 times greater than with the cavity dump. The increase in signal and signal to noise ratio associated with this increase of measurement time allowed us to distinguish nearly as many sidebands with 5 kV-cm$^{-1}$ applied as were observed with 11 kV-cm$^{-1}$ applied. The grey crosshatch pattern in Fig. 3b again represents the confidence limit for identifying sidebands.

Our observation of high-order sidebands in bulk GaAs shows that quantum confinement is not necessary to observe high-order sideband generation. We speculate that HSG is observable in any direct-gap semiconductor. Studies of HSG in different materials systems should lead to not only an improved understanding of electron-hole recollisions but also to a better understanding of the properties of excitons in these materials. Additionally, the persistence of HSG to temperatures greater than 170 K may allow for experiments to be performed cryogen-free, which would be particularly beneficial for developing practical technology based on this phenomenon.

The authors would like to thank D. Enyeart for his work in operating the free electron laser during the experiments performed. This work was supported by NSF-DMR grant 1006603.

[1] M. Ferray, A. L'Huillier, X. F. Li, L. A. Lompre, G. Mainfray, and C. Manus, J. Phys. B **21** (3), L31 (1988).

[2] J. L. Krause, K. J. Schafer, and K. C. Kulander, Phys. Rev. Lett. **68** (24), 3535 (1992); P. B. Corkum, Phys. Rev. Lett. **71** (13), 1994 (1993); F. Krausz and M. Ivanov, Rev. Mod. Phys. **81** (1), 163 (2009).

[3] P. B. Corkum and F. Krausz, Nat. Phys. **3** (6), 381 (2007).

[4] J. Itatani, J. Levesque, D. Zeidler, H. Niikura, H. Pepin, J. C. Kieffer, P. B. Corkum, and D. M. Villeneuve, Nature **432** (7019), 867 (2004); C. Vozzi, M. Negro, F. Calegari, G. Sansone, M. Nisoli, S. De Silvestri, and S. Stagira, Nat. Phys. **7** (10), 822 (2011).

[5] S. Ghimire, A. D. DiChiara, E. Sistrunk, P. Agostini, L. F. DiMauro, and D. A. Reis, Nat. Phys. **7** (2), 138 (2011).

[6] W. Franz, Z. Naturforsch. A **13A**, 484 (1958); L. V. Keldysh, Sov. Phys. JETP **7**, 788 (1958); K. B. Nordstrom, K. Johnsen, S. J. Allen, A. P. Jauho, B. Birnir, J. Kono, T. Noda, H. Akiyama, and H. Sakaki, Phys. Rev. Lett. **81** (2), 457 (1998).

[7] S. G. Carter, V. Birkedal, C. S. Wang, L. A. Coldren, A. V. Maslov, D. S. Citrin, and M. S. Sherwin, Science **310** (5748), 651 (2005); M. Wagner, H. Schneider, D. Stehr, S. Winnerl, A. M. Andrews, S. Schartner, G. Strasser, and M. Helm, Phys. Rev. Lett. **105** (16), 167401 (2010); B. Zaks, D. Stehr, T. A. Truong, P. M. Petroff, S. Hughes, and M. S. Sherwin, New J. Phys. **13** (8), 083009 (2011).

[8] J. Černe, J. Kono, T. Inoshita, M. Sherwin, M. Sundaram, and A. C. Gossard, Appl. Phys. Lett. **70** (26), 3543 (1997).

[9] J. Kono, M. Y. Su, T. Inoshita, T. Noda, M. S. Sherwin, S. J. Allen, and H. Sakaki, Phys. Rev. Lett. **79** (9), 1758 (1997); M. Wagner, H. Schneider, S. Winnerl, M. Helm, T. Roch, A. M. Andrews, S. Schartner, and G. Strasser, Appl. Phys. Lett. **94** (24), 241105 (2009).

[10] B. Zaks, R. B. Liu, and M. S. Sherwin, Nature **483** (7391), 580 (2012).

[11] R. B. Liu and B. F. Zhu, in *Proceedings of the 28th International Conference on the Physics of Semiconductors*, Vienna, Austria, 24-28 July 2006, edited by W. Jantsch and F Schäffler (American Institute of Physics, 2007), pp. 1455.

[12] J. Y. Yan, Phys. Rev. B **78** (7), 075204 (2008).

[13] J. P. Kaminski, J. S. Spector, C. L. Felix, D. P. Enyeart, D. T. White, and G. Ramian, Appl. Phys. Lett. **57** (26), 2770 (1990).